\documentstyle[12pt,psfig,aasms4]{article}
\def\degr{\hbox{$^\circ$}}
\def\arcmin{\hbox{$^\prime$}}
\def\arcsec{\hbox{$^{\prime\prime}$}}

\def\fs{\hbox{$.\!\!^{\rm s}$}}

\def\farcm{\hbox{$.\mkern-4mu^\prime$}}
\def\farcs{\hbox{$.\!\!^{\prime\prime}$}}         
\def\gsim{\mathrel{\hbox{\rlap{\lower.55ex \hbox {$\sim$}}
                   \kern-.3em \raise.4ex \hbox{$>$}}}}
\begin{document}  
%\psdraft
\title{A Search for Optical Afterglow from GRB970828}
\author{P.J. Groot\altaffilmark{1}, T.J. Galama\altaffilmark{1},
J. van Paradijs\altaffilmark{1,2}, C. Kouveliotou\altaffilmark{3,4},
R.A.M.J. Wijers\altaffilmark{5}, 
J. Bloom\altaffilmark{5,6},
N. Tanvir\altaffilmark{5}, 
R. Vanderspek\altaffilmark{7},
J. Greiner\altaffilmark{8}, 
A.J. Castro-Tirado\altaffilmark{9}, 
J. Gorosabel\altaffilmark{9},
T. von Hippel\altaffilmark{10}, 
M. Lehnert\altaffilmark{11},
K. Kuijken\altaffilmark{12},
H. Hoekstra\altaffilmark{12}, N. Metcalfe\altaffilmark{13},
C. Howk\altaffilmark{10}, C. Conselice\altaffilmark{10}
J. Telting\altaffilmark{14}, R.G.M. Rutten\altaffilmark{14},
J. Rhoads\altaffilmark{15}, 
A. Cole\altaffilmark{10}, 
D.J. Pisano\altaffilmark{10}, R. Naber\altaffilmark{12}, 
R. Schwarz\altaffilmark{8}}

\altaffiltext{1}{Astronomical Institute `Anton Pannekoek', University
of Amsterdam,
\& Center for High Energy Astrophysics,
Kruislaan 403, 1098 SJ Amsterdam, The Netherlands}
\altaffiltext{2}{Physics Department, University of Alabama in
Huntsville, Huntsville AL 35899, USA}
\altaffiltext{3}{Universities Space Research Asociation at NASA/MSFC}
\altaffiltext{4}{NASA/MSFC, Code ES-84, Huntsville AL 35812, USA}
\altaffiltext{5}{Institute of Astronomy, Madingley Road, Cambridge, UK}
\altaffiltext{6}{Division of Physics, Mathematics and Astronomy
105-24, Caltech, Pasadena CA 91125}
\altaffiltext{7}{Center for Space Research, MIT, Cambridge MA 02139, USA}
\altaffiltext{8}{Astrophysikalisches Institut Postdam, Germany}
\altaffiltext{9}{ Laboratorio de Astrof\'{\i}sica Espacial y F\'{\i}sica
Fundamental (LAEFF-INTA), P.O. Box 50727, E-28080, Madrid, Spain}
\altaffiltext{10}{University of Wisconsin, Astronomy Department, Madison, 
WI 53706, USA}
\altaffiltext{11}{Sterrewacht Leiden, Postbus 9513, 2300RA Leiden, The Netherlands}
\altaffiltext{12}{Kapteyn Astronomical Institute, Postbus 800, 9700 AV,
Groningen, The Netherlands} 
\altaffiltext{13}{Physics Department, University of Durham, South Road, 
Durham, UK}
\altaffiltext{14}{ING Telescopes/NFRA, Apartado 321, Sta. Cruz de La
Palma, Tenerife 38780, Spain} 
\altaffiltext{15}{NOAO, Tucson AZ 85726, USA}

\begin{abstract}
We report on the results of $R$ band observations of the error box of the 
$\gamma$-ray burst of August 28, 1997, made between 4 hours and 
8 days after this burst occurred. No counterpart was found varying by more 
than 0.2 magnitudes down to $R = 23.8$. We discuss the consequences of this 
non-detection for relativistic blast wave models of $\gamma$-ray
bursts, and the possible effect of redshift on the relation between
optical absorption and the low-energy cut off in the X-ray afterglow spectrum.
\end{abstract}


\section{Introduction}
Since the first discovery of a $\gamma$-ray burst (GRB) in 1967 
(Klebesabel et al., 1973) these short outbursts of highly energetic 
photons have formed one of astronomy's most elusive problems. Following
the discovery by Meegan et al. (1992) of their isotropic sky
distribution and inhomogeneous spatial distribution (which excluded
that GRBs originate from a galactic-disk source population) the
discussion on the nature of GRB sources focussed on their distances:
either of order 10$^5$ pc ('galactic halo' model), or several Gpc
('cosmological' model). The association  of the optical counterpart of
GRB970228 (Groot et al., 1997; Van Paradijs et al., 1997) with what is
most likely a galaxy (Groot et al, 1997b; Metzger et al., 1997a; Sahu
et al., 1997) and especially the determination of a
redshift for GRB970508 (Metzger et al., 1997b) have shown that GRBs are
located at cosmological distances, and are thereby the most luminous photon 
sources known in the Universe. 
The question of what causes GRBs has now become the
centerpiece of the discussion, and the detection of more optical 
counterparts is a key element in determining this cause. 

In this Letter we report on our search for a transient optical 
counterpart for GRB970828, based on observations made with the 4.2m William
Herschel Telescope (WHT) on La Palma, and the 3.5m WIYN Telescope on 
Kitt Peak. None was detected, down to a magnitude level $R = 23.8$.

GRB\,970828 was discovered with the All-Sky Monitor 
(ASM) on the Rossi X-ray Timing Explorer (RXTE) on August 28, 1997, 
UT $17^{\rm h}44^{\rm m}36^{\rm s}$ 
from an elliptical region centered at RA=18$^{\rm h}$08$^{\rm m}$29$^{\rm s}$, 
Dec=+59\degr18\farcm0 (J2000), with a major axis of 5\farcm0, and a
minor axis of 2\farcm0 (Remillard et al. 1997; Smith et al. 1997). 
Within 3.6 hours the RXTE/PCA scanned the region of the sky around the error 
box of the ASM burst, and detected a weak X-ray source, 
located in the ASM error box with a 2--10 keV flux of 0.5 mCrab 
(Marshall et al. 1997). 
The burst was also detected with the Burst And Transient Source Experiment 
BATSE and the GRB 
experiment on Ulysses. Its fluence and peak flux were 
$7 \times 10^{-5}$ erg cm$^{-2}$, and $3 \times 10^{-6}$ erg 
cm$^{-2}$ s$^{-1}$, respectively. From 
the difference between burst arrival times, its position was
constrained to lie within 
a 1.62 arcminute wide annulus, that intersected the RXTE error box
(Hurley et al. 1997). In 
an ASCA observation made between Aug. 29.91 and 30.85 UT, a weak X-ray 
source was detected at an average flux level of $4\times 10^{-13}$
erg~cm$^{-2}$~s$^{-1}$ (2--10 keV). The ASCA error box is centered on 
RA=18$^{\rm h}$08$^{\rm m}$32\fs3, 
Dec =+59\degr18\arcmin54\arcsec (J2000) and has a 0\farcm5 radius 
(Murakami et al. 1997).

\section{Observations and Data Analysis}

We observed the GRB error box with the Prime Focus Camera of the 
WHT, on 9 nights between August 28, UT 21$^{\rm h}$47$^{\rm m}$, 
and September 5, UT 22$^{\rm h}$07$^{\rm m}$ (see Table 1). 
The first observation was made just over 4 hours after the
$\gamma$-ray burst. All observations were made with a Cousins $R$ 
band filter (Bessell 1979). 
During the first two nights and the last three nights, we used a LORAL 
2048$\times$2048 CCD chip, with 15$\mu$ pixels, giving a field of view of 
8\farcm45$\times$8\farcm45. During the intervening nights 
we used an EEV CCD chip 
(2048$\times$4096), windowed at 2048$\times$2400, 
with 13.5$\mu$ pixels giving a 8\farcm1$\times$9\farcm5 field of view. 
On August 30 two R band images were made with the WIYN Telescope. The 
camera contained a $2048\times 2048$ CCD, giving a field of view of 
$6\farcm8\times 6\farcm8$. 

We obtained a photometric calibration of the CCD images from observations 
of Landolt Selected Area 113, stars 281, 158, 183 and 167 (Landolt
1992), on Aug 31, $0^{\rm h}14^{\rm m}$ UT with the WHT. 

A region of 2\arcmin$\times$2\arcmin ~centered on the ASCA position 
in the bias-subtracted and flatfielded images was analyzed using 
DoPhot (Schechter et al., 1993), in which astrometric and photometric 
information of all objects are determined from bivariate Gaussian function 
fits to the brightness distribution in their image; the 
parameters of these fits also tell us whether an object is stellar (i.e., 
unresolved) or a galaxy.  
In this region (see Fig. 1) we find a total of 63 objects, 36 of which 
are stellar, and 27 galaxies, down to $R$=23.8.  

We have searched for variable objects by comparing 
the magnitudes of each star as determined for each of the images. 
Comparison of images taken on different nights showed no variation on time 
scales between a day and a week in excess of 
0.2 mag for $R\leq$23.8 (for the last three nights the limit on variability 
is 0.3 mag for $R\leq$23.8). Comparison of three images taken on the night 
of August 29 to 30 showed no variations on time scales of several hours in 
excess of 0.2 mag for $R< 22.5$.

\section{Discussion}

\subsection{Comparison with optical afterglows of GRB970228 and GRB970508}

The large variation in optical response of GRBs (relative to their strength 
in $\gamma$ rays) was already clear from a comparison between GRB970228 and 
GRB970111. Within a day after GRB970228 ocurred it showed an optical afterglow 
at $R = 20.8$ (Van Paradijs et al. 1997; Galama et al. 1997a; Pedichini et al. 
1997; Guarneri et al. 1997). GRB970111 was not detected in optical 
observations
made 19 hours after it occurred ($R>20.8$, and $R>22.6$, for variations in 
excess of 0.2 and 0.5 magnitudes, respectively, Castro-Tirado et al. 1997), 
in spite of the fact that its $\gamma$-ray fluence (Galama et al.,
1997b) was five times larger than that of GRB970228 (Costa et al.,
1997). Since only one deep image was made in the week 
following GRB970111, its non-detection may have been the 
result of, e.g., a very rapid decay of any optical afterglow, or a very 
slow rise thereof (like for GRB970508, see Bond 1997; Djorgovski et al. 
1997; Sahu et al. 1997b; Galama et al. 1997c).

The non-detection ($R>23.8$ for variations in excess of 0.2 magnitudes) 
of GRB970828 during our optical observations, which 
covered the time interval between 4 hours and 8 days after the 
burst at intervals of a day, 
show the very large range in optical responses of GRBs in an even 
more striking fashion. We have used the fluence, $E_{\rm GRB}$ (in
ergs cm$^{-2}$), as a measure 
of the GRB strength, and compared the ratio of the optical 
peak flux to the GRB fluence of GRB970828 with that of GRB970508. 
The latter had a peak magnitude $R=19.8$ (Mignoli et al., 1997), 
therefore the difference in
optical peak luminosities between GRB970508 and GRB970228 is more than
4 magnitudes. The ratio of their fluences, 
$E_{\rm GRB}(970828)/E_{\rm GRB}(970508) =24$ (Kouveliotou et al. 1997a,b). 
Thus, we find that the optical peak response of GRB970828, with
respect to its $\gamma$-ray fluence, is a factor $\sim 10^3$
smaller than that of GRB970508. (Compared to GRB970228 the difference is
a factor $>10^2$.)

We have made a similar 
comparison with published X-ray afterglow fluxes ($F_{\rm X}$) for the two 
GRBs with optical afterglow. 
Most of these refer 
to the energy range 2--10 keV. Only the ROSAT fluxes had to be
transformed to this range; in doing this we assumed a power law X-ray
spectrum with photon index in the range --1.4 to --2.0 (Costa et
al. 1997; Yoshida et al. 1997). 
This range leads to an uncertainty in the transformed ROSAT flux of
less than a factor 2. 
 The results, in the form of the ratio 
${\cal R}_{\rm X} = F_{\rm X}/E_{\rm GRB}$, are summarized in Fig. 2, 
which shows 
the variation of this quantity as a function of the time interval since 
the burst, for four bursts with published X-ray afterglow
information. 
This figure shows that the differences in ${\cal R}_{\rm X}$ between
these bursts 
are moderate (less than an order of magnitude). It is noteworthy that the two 
bursts with optical counterparts also have the highest values of 
${\cal R}_{\rm X}$ (for a given value of $\Delta t$).

We finally compared the peak flux in the $R$-band afterglows with the 
brightness of the X-ray afterglow. In view of their rather similar decay 
rates we used for the latter the 2--10 keV flux 
as measured 1 day 
after the GRB occurred, $F_{\rm X}$(1 day). The corresponding ratio 
$F_{\rm peak}(R {\rm band})/F_{\rm X}$(1 day) for GRB970828 differs by a 
factor
$> 150$ from that for GRB970508, and a factor $>10$ from that for GRB970228.

\subsection{Comparison with Relativistic Blast Wave Models}

A relatively succesful way of explaining the existence of GRB afterglows 
(at all wavelengths) has been the so-called blastwave or fireball
models (e.g. M\'esz\'aros 1995). These
models involve the generation of a massive amount of energy in a very
small, compact region, by an unexplained mechanism. The result of this
dumping of energy is a relativistically expanding fireball
(blastwave), that collides with the interstellar or circumstellar
medium and generates shocks that emit the synchrotron radiation that
is observed as the afterglow.  

Figure 3 shows the
available data for GRB970828 in $\gamma$-rays, X-rays, $B$, and $R$,
plus simple blastwave model fits, which are normalized to agree with the X-ray
data. If we compare this with the data available for GRB970228 (Wijers
, Rees and M\'esz\'aros, 1997), it is striking that the decay part of the
X-ray curves  are
virtually the same for these two bursts (i.e. in slope {\em and}
offset).  
But whereas the first stages of the optical decay for GRB\,970228 are in
good agreement with the afterglow prediction (Wijers, Rees, and
M\'esz\'aros 1997), the
earliest upper limit to the optical brightness of GRB\,970828 is 300 times
lower than the predicted value. 

The simplest spherically symmetric blastwave models for GRB afterglows         
require that the slope of the spectrum follows from the slope of     
the temporal decay, once the decay curve is measured in one wavelength
band and is found to be a pure power law. 
From that, the offset in brightness at any other waveband  
is fixed and the predicted flux at that waveband is hard to change.            
M\'esz\'aros, Rees, \& Wijers (1997) showed that if the blastwave
is beamed one can get different relations between spectral and temporal        
slopes, giving possibly much smaller offsets between the optical and X-ray     
light curves of the afterglow. As an example, let the energy per unit          
solid angle, $E$, vary with angle from the jet axis, $\theta$, as              
$E\propto\theta^{-4}$ and the Lorentz factor $\Gamma\propto\theta^{-1}$.       
Then a temporal decay rate $F\propto t^{-1.3}$ as seen here would              
occur for a spectrum $F_\nu\propto\nu^{0.4}$, i.e.\ it would rise from         
optical to X rays and the predicted $R$ band curve would be a factor 24        
below the X-ray curve. At the time of our first limit this model would give    
$R=28.4$, quite consistent with the data.                                      
                                                                               
\subsection{Absorption in Redshifted Material}

Another explanation, pointed out to us by
dr. B. Paczy\'nski, for the non-detection of optical     
afterglow could be photoelectric absorption, also visible as a low-energy
cut-off in the X-ray spectrum. If we assume a modest hydrogen column
density of $N_{\rm H}$ $\sim 10^{21}$ atoms cm$^{-2}$ and make the
assumption that the absorbing material is at redshift $z$=0, this
would imply 0.34 magnitudes of extinction in the R
band (Gorenstein 1975; Cardelli et al. 1989). 

In case the absorption takes place at some redshift $z$ the effect is
a bit more complicated. 
The cross section for photo-electric absorption in the (0.2--5) keV
range depends on energy 
roughly as $E^{-2.6}$ (Morrison and McCammon 1983). Then the factor by 
which the apparent $N_{\rm H}$, inferred from the low-energy cut-off
in the X-ray spectrum, has to be increased is approximately 
$(1+z)^{2.6}$. If we assume, for example, that the GRB occured at a
redshift of $z$=1, the factor by which the apparent value of $N_{\rm
H}$ has to be increased would be $\sim$6. 
Moreover, the photons in the $R$ band we observe would be at wavelengths near 
3200 \AA\ at the source, at which wavelength the interstellar 
absorption is approximately a factor 2.5 larger than in the $R$ band 
(Cardelli et al., 1989). These
combined effects would lead, for a GRB at $z$=1 and an apparent,
moderate, $N_{\rm H}$=10$^{21}$ atoms cm$^{-2}$ to an $R$ band 
extinction of $\sim$5 mags.

If absorption is the correct explanation, a substantial fraction 
of GRB sources (those with a very small optical response) 
would be located close to where large column densities are 
available, i.e., in disks of galaxies. This would link GRBs to a population 
of massive stars. This is expected for the failed-supernova model and for 
the hypernova model, proposed by Woosley (1993) and Paczy\'nski
(1997), respectively.
In view of the large kick velocities imparted on neutron stars at birth 
(Lyne \& Lorimer 1994; Hansen \& Phinney 1997; Van den Heuvel \& Van Paradijs 
1997) it remains to be seen whether a merging neutron star binary model 
would be consistent with this consequence.

\vspace{1cm}
{\bf Acknowledgments} We thank the RXTE ASM and PCA teams for their very 
fast response to and communications regarding the $\gamma$-ray burst of 
August 28, 1997. We thank B. Paczy\'nski and W. Lewin for enlightening 
discussions on the importance of redshift for absorption of optical 
afterglows. TG is supported by NFRA under grant no. 781.76.011.
JG and RS are supported by the Deutsche Agentur f\"ur
Raumfahrtangelegenheiten (DARA) GmbH under contract FKZ 50 QQ 9602 3 and
50 OR 9206 8, respectively.

\section{References}

Bessel, M.S., 1979, PASP 91, 589

Bond, H.E. 1997, IAU Circular 6654

Castro-Tirado, A. et al. 1997, IAU Circular 6598

Cordelli, J.A, Clayton, G.C, Mathis, J.S., 1989, ApJ 345, 245

Costa, E. et al. 1997, Nature 387, 783

Djorgovski, S.G. et al. 1997, Nature 387, 876

Frontera, F. et al. 1997, IAU Circular 6637

Galama, T.J. et al. 1997a, Nature 387, 479

Galama, T.J. et al., 1997b, ApJ 486, L5

Galama, T.J. et al. 1997c, ApJ (in preparation)

Gorenstein, P., 1975, ApJ 198, 95

Groot, P.J. et al., 1997a, IAU Circular 6584

Groot, P.J. et al., 1997b, IAU Circular 6588

Guarneri, A. et al. 1997, private communication

Hansen, B.M.S, Phinney, E.S., 1997, MNRAS, in press

Hurley, K. et al. 1997, IAU Circular 6728

Klebesabel, R., et al., 1973, ApJ 182, L85

Kouveliotou, C. et al., 1997a, IAU Circular 6660

Kouveliotou, C. et al., 1997b, private communication

Landolt, A. 1992, AJ 104, 340

Laureijs, R., 1989, PhD thesis, Groningen University

Lyne, A., Lorimer, 1994, Nature 269, 127

Marshall, F.A. et al., 1997, IAU Circular 6727

Meegan, C.A. et al., 1992, Nature 355 143
 
M\'esz\'aros, P., 1995, Gamma-ray burst models: General requirements and
predictions, in Proc. 17th Texas Symposium on Relativistic Astrophysics, Annals
N.Y. Acad. Sci., bf 759, 440-445

M\'esz\'aros, Rees, \& Wijers, 1997, ApJ, submitted(astro-ph/9709273)     

Metzger, M. et al., 1997a, IAU Circular 6588

Metzger, M. et al., 1997b, Nature 387, 878
 
Mignoli, M. et al., 1997, IAU Circular 6661

Morrison, R. and McCammon, D., 1983, ApJ 270, 119
 
Murakami, T. et al., 1997, IAU Circular 6732

Paczy\'nski, B. 1997, ApJ (in press)

Pedichini et al. 1997, A\&A (in press)

Piro, L. et al. 1997, IAU Circular 6656

Remillard, R.A. et al. 1997, IAU Circular 6726

Sahu, K. et al., 1997a, 387, 476

Sahu, K. et al. 1997b, ApJ (in press)

Schechter, P.L., Mateo., M., and Saha, A., 1993, PASP 105, 1342

Smith, D. et al. 1997, IAU Circular 6728

Van den Heuvel, E.P.J., Van Paradijs, J., 1997, ApJ 483, 399

Van Paradijs, J. et al. 1997, Nature 386, 686

Wijers, R.A.M.J., Rees, M.J. \& Meszaros, P. 1997, MNRAS 288, L51

Woosley, S.E., 1993, ApJ 405, 273

Yoshida, A. et al. 1997, IAUC 6593

\newpage

\begin{table}
\caption[]{Log of observations GRB970828}
\begin{tabular}{lllll}
Date 	& Telescope & UT Start & Exp. time (s) & Seeing\\ \hline\\[2mm]
Aug 28 	& WHT & 21:47    & 900 	       & 0\farcs86\\
Aug 29  & WHT & 21:15    & 900	       & 0\farcs74\\
Aug 30  & WIYN & 05:08    & 600               & 0\farcs8\\
Aug 30  & WIYN & 07:38    & 900               & 1\farcs2\\
Aug 30  & WHT & 23:22	   & 900	       & 0\farcs90\\
Aug 31  & WHT & 20:54	   & 900	       & 0\farcs71\\
Sep  1  & WHT & 21:16	   & 600	       & 0\farcs80\\
Sep  2  & WHT & 20:53	   & 600	       & 0\farcs76\\
Sep  3  & WHT & 22:44	   & 600	       & 0\farcs88\\
Sep  4  & WHT & 21:53	   & 600	       & 0\farcs79\\
Sep  5  & WHT & 22:07    & 600               & 0\farcs86\\
\end{tabular}
\end{table}

\newpage

Figure Captions

Fig. 1: 2\arcmin$\times$2\arcmin $R$-band image of the sky region 
centered on the $0.5^{\prime}$ 
radius ASCA error box of GRB970828, taken at the WHT on Sept 2. 

Fig. 2: Variation of the ratio $R_{\rm X}$ of the afterglow (2-10 keV) 
X-ray flux (see text) to the fluence in the $\gamma$-ray burst as
function of time in days. GRB 
fluences were obtained from the following sources. GRB970228: Costa et al. 
(1997); GRB970508: Kouveliotou et al. (1997a); 
GRB970828: Kouveliotou et al. (1997b). The X-ray fluxes were 
obtained from the following sources: GRB970228: Costa et al. (1997), 
Yoshida et al. (1997), Frontera et al. (1997); GRB970508: Piro et al. 
(1997); GRB970828: Marshall et al. (1997), Murakami et al. 
(1997b).

Fig. 3: Variation of the observed fluxes in $\gamma$ rays, X rays, and in 
the $B$ and $R$ bands of GRB970828, together with simple blast wave model
fits as described in Wijers, Rees and M\'esz\'aros (1997).  

\newpage

\begin{figure}
\centerline{\psfig{figure=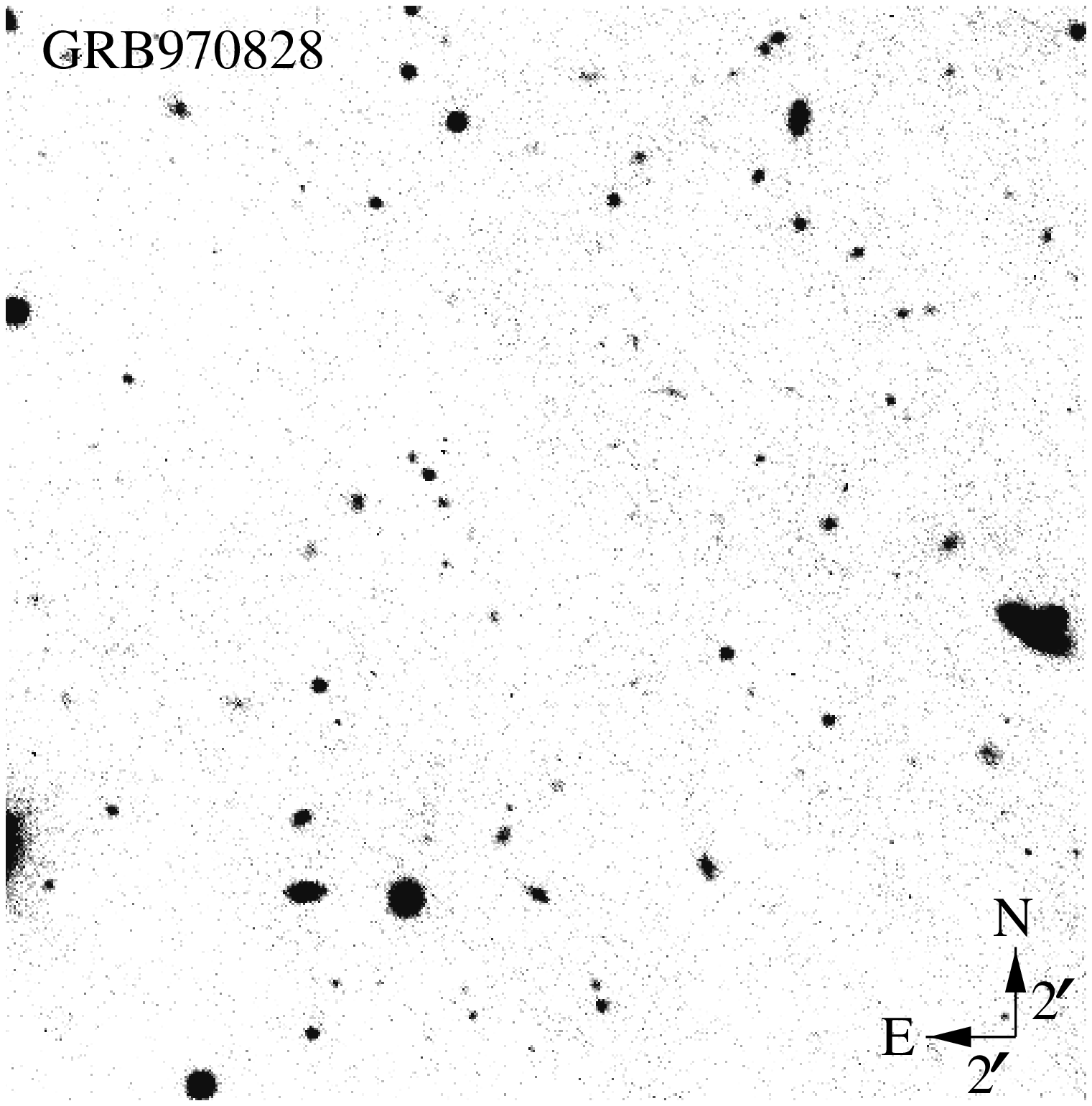,width=12cm,angle=-90}}
\end{figure}

\begin{figure}
\centerline{\psfig{figure=FxEbt.ps,angle=-90,width=12cm,clip=}}
\end{figure}

\begin{figure}
\centerline{\psfig{figure=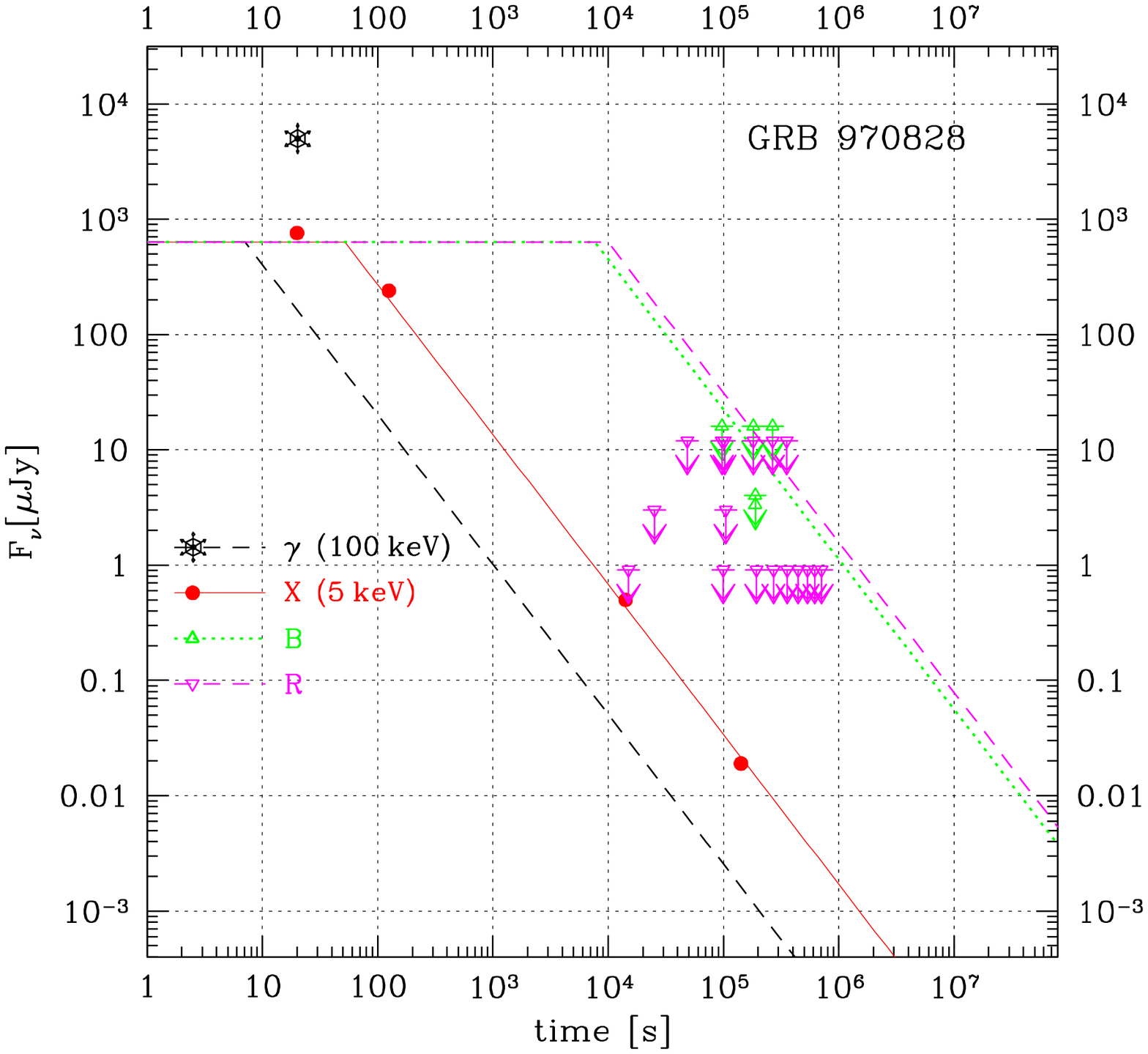,width=12cm}}
\end{figure}

\end{document}